\documentclass[]{aa}

\usepackage{natbib}
\usepackage{graphicx}
\usepackage{txfonts}
\usepackage{gensymb}
\usepackage{amsmath}

\usepackage{placeins}

\usepackage[switch]{lineno}
\bibpunct{(}{)}{;}{a}{,}{,}

\usepackage{pstricks}
\usepackage{color}

\usepackage{hyperref}
\hypersetup{
    bookmarks=true,         % show bookmarks bar?
    unicode=true,           % non-Latin characters in Acrobat's bookmarks
    pdftoolbar=true,        % show Acrobat's toolbar?
    pdfmenubar=true,        % show Acrobat's menu?
    pdffitwindow=true,      % page fit to window when opened
    pdfauthor={Marsset et al.},
    pdfsubject={Planetary Science},
    pdfkeywords={},         % list of keywords
    pdfnewwindow=true,      % links in new window
    colorlinks=true,        % false: boxed links; true: colored links
    linkcolor=gray,         % color of internal links
    citecolor=blue,         % color of links to bibliography
    filecolor=gray,         % color of file links
    urlcolor=gray           % color of external links
}

\newcommand{\numb}[1]{\textcolor{green}{#1}}

\renewcommand{\numb}[1]{#1}  %%% comment to highlight numbers

\begin{document} 
\bibliographystyle{aa}

  \title{3D shape of asteroid (6)~Hebe from VLT/SPHERE
    imaging: Implications for the origin of ordinary H
    chondrites\thanks{Based on observations made with ESO Telescopes
      at the La Silla Paranal 
      Observatory under programme ID 60.A-9379 and 086.C-0785.}
  }
   
  \author{M. Marsset\inst{1}
    \and
    B. Carry\inst{2,3}
    \and
    C. Dumas\inst{4}
    \and
    J. Hanu\v s\inst{5}
    \and
    M. Viikinkoski\inst{6}
    \and
    P. Vernazza\inst{7} 
    \and
    T. G. M\"uller\inst{8}
    \and
    M. Delbo\inst{2}
    \and
    E. Jehin\inst{9}
    \and
    M. Gillon\inst{9}
    \and
    J.Grice\inst{2,10}
    \and
    B. Yang\inst{11}
    \and
    T. Fusco\inst{7,12}
    \and
    J. Berthier\inst{3}
    \and
    S. Sonnett\inst{13}
    \and
    F. Kugel\inst{14}
    \and
    J. Caron\inst{14}
    \and
    R. Behrend\inst{14}
  }

  \institute{Astrophysics Research Centre, Queen's University Belfast, BT7 1NN, UK \\
    \email{michael.marsset@qub.ac.uk}
    \and            
    Universit\'e C{\^o}te d'Azur, Observatoire de la C{\^o}te d'Azur, Lagrange, CNRS, France
    \and
    IMCCE, Observatoire de Paris, PSL Research University, CNRS, Sorbonne Universit\'es, UPMC Univ Paris 06, Univ. Lille, France
    \and  
    TMT Observatory, 100 W. Walnut Street, Suite 300, Pasadena, CA 91124, USA
    \and
     Astronomical Institute, Faculty of Mathematics and Physics, Charles University, V Hole\v sovi\v ck\' ach 2, 18000 Prague, Czech Republic
    \and
    Department of Mathematics, Tampere University of Technology, PO Box 553, 33101 Tampere, Finland
    \and
    Aix Marseille Univ, CNRS, LAM, Laboratoire d'Astrophysique de Marseille, Marseille, France
    \and    
    Max-Planck-Institut f\"ur extraterrestrische Physik, Giessenbachstrasse, 85748 Garching, Germany
    \and
    Space sciences, Technologies and Astrophysics Research Institute, Universit\'e de Li\`ege, All\'ee du 6 Ao\^ut 17, 4000 Li\`ege, Belgium
    \and
    Open University, School of Physical Sciences, The Open University, MK7 6AA, UK
    \and
    European Southern Observatory (ESO), Alonso de C\'ordova 3107, 1900 Casilla Vitacura, Santiago, Chile
    \and
    ONERA - the French Aerospace Lab, F-92322 Ch\^atillon, France
    \and
    Planetary Science Institute, 1700 East Fort Lowell, Suite 106, Tucson, AZ 85719, USA
    \and
    CdR \& CdL Group: Lightcurves of Minor Planets and Variable Stars,
    Observatoire de Gen{\`e}ve, CH-1290 Sauverny, Switzerland
  }
  
  \date{Received XXX; accepted XXX}

  \abstract
      {
        The high-angular-resolution capability of the new-generation ground-based
        adaptive-optics camera
        SPHERE at ESO VLT allows us to assess, for the very first time, the
        cratering record 
        of medium-sized (D$\sim$100-200\,km) asteroids from the ground, opening 
        the prospect of a new era of investigation of the asteroid
        belt's collisional history.
      }
      {
        We investigate here the collisional history of asteroid (6) Hebe 
        and challenge the idea that Hebe may be the parent body of
        ordinary H chondrites, the most common type of meteorites 
        found on Earth ($\sim$34\% of the falls).
      }   
      {
        We observed Hebe with SPHERE as part of the science verification of the instrument.
        Combined with earlier adaptive-optics images and
        optical light curves, we model the
        spin and three-dimensional (3D) shape of Hebe 
        and check the consistency of the derived model against available
        stellar occultations and thermal measurements.
        %Hebe's surface properties were further derived through thermophysical modelling.
      }
      {
        Our 3D shape model fits the images with sub-pixel residuals and
        the light curves to \numb{0.02} mag.
        The rotation period (\numb{7.274\,47}\,h),
        spin (ECJ2000 $\lambda$,$\beta$ of \numb{343\degr,+47\degr}), and
        volume-equivalent diameter (\numb{$193\pm6$}\,km) are
        consistent with previous determinations and thermophysical modeling.
        %The later further indicate a thermal inertia $\Gamma$ of $\sim$50~J\,m$^{-2}$\,s$^{-0.5}$K\,$^{-1}$,
        %high surface roughness and regolith grain size of 0.2--0.3\,mm.
        Hebe's inferred density is
        \numb{$3.48\pm0.64$}~g.cm$^{-3}$, 
        in agreement with an intact interior 
        based on its H-chondrite composition. Using the 3D shape model
        to derive the volume of the largest depression (likely impact
        crater), it appears that the latter is  
        significantly smaller than the total volume of close-by S-type
        H-chondrite-like asteroid families. 
      }
      {
        Our results imply that (6)~Hebe is not the most likely source of
        H~chondrites. Over the coming years, our team will collect similar high-precision shape
        measurements with VLT/SPHERE for $\sim$40 asteroids covering the main compositional classes,
        thus providing an unprecedented dataset to investigate the origin
        and collisional evolution of the asteroid belt.

      }


  \titlerunning{Origin of H chondrites from 3D shape modelling of (6) Hebe}
  \authorrunning{Marsset et al.}

  \maketitle

\section{Introduction}
\label{sec:introduction}

  \indent Disk-resolved imaging is a powerful tool to investigate the origin and collisional history 
  of asteroids. This has been remarkably illustrated by fly-by and rendezvous space missions
  \citep{Belton:1992jq, Belton:1996ka, Zuber:2000km, 
  Fujiwara:2006ca, Sierks:2011il, Russell:2012ep, Russell:2016dj}, as well as observations from the Earth
  (e.g., \citealt{Carry:2008ip, Carry:2010je, Merline:2013fs}). 
  In the late nineties, observations of (4)~Vesta with the Hubble
  Space Telescope (HST) led to the discovery 
  of the now-called ``Rheasilvia basin''  and allowed for establishment of the origin of
  the Vestoids and HED meteorites found on Earth \citep{ThomasPC:1997cu, Binzel:1997bj}. 
  Specifically, it was demonstrated that the basin-forming event on Vesta excavated enough material  
  to account for the family of small asteroids with spectral properties similar to Vesta.
  HST observations thus confirmed the origin of these bodies as fragments from Vesta, as 
  previously suspected based on spectroscopic measurements \citep{Binzel:1993ju}.
  Recently, the Rheasilvia basin was revealed in much greater detail by the Dawn mission, which 
  unveiled two overlapping giant impact features \citep{Schenk:2012ina}. 
  
 % Due to the limited resolving power of available instruments, 
  % disk-resolved Earth based observations long remained limited to the 
  %largest objects like Ceres and Vesta (e.g., \citealt{ThomasPC:1997cu, Carry:2008ip}). 
  %As a result, the current dataset of disk-resolved images is mainly composed of in-situ
 % measurements obtained via fly-by and rendezvous missions
  %\citep{Belton:1992jq, Belton:1996ka, Zuber:2000km, 
  %Fujiwara:2006ca, Sierks:2011il, Russell:2012ep, Russell:2016dj}.
    
  \indent In the 2000's, a new generation of ground-based imagers with 
  high-angular-resolution capability, such as NIRC2
  \citep{Wizinowich:2000ua, vanDam:2004jd} on the W. M. Keck II
  telescope and NACO \citep{Lenzen:2003iu, Rousset:2003hh} on the
  European Southern Observatory (ESO) Very Large Telescope (VLT),
  made disk-resolved imaging achievable from the ground for 
  a larger number of medium-sized ($\sim$100-200-km in diameter) asteroids.
  In turn, these observations triggered the development of methods for 
  modeling the tridimensional shape of these objects by combining the images 
  with optical light curves
  (see, e.g., \citealt{Carry:2010bo, Carry:2012jn, Kaasalainen:2011tk, Viikinkoski:2015jha}). 
  These models were subsequently validated by in-situ measurements performed by the 
  ESA Rosetta mission during the fly-by of asteroid (21)~Lutetia 
  \citep{Sierks:2011il, Carry:2010je, Carry:2012jn, ORourke:2012ch}.
  
  More recently, the newly commissioned VLT/Spectro-Polarimetric High-contrast Exoplanet 
  Research instrument (SPHERE) and its
  very high performance adaptive optics system
  \citep{Beuzit:2008gt} demonstrated its ability to reveal in even greater detail
  the surface of medium-sized asteroids by resolving their largest (D$>$30km) 
  craters \citep{Viikinkoski:2015jk, Hanus:2017ed}. This remarkable achievement
  opens the prospect of a new era of exploration of the asteroid belt and its collisional history.
  
  \indent Here, we use VLT/SPHERE to investigate the shape and topography of 
  asteroid (6)~Hebe, a large main-belt asteroid \citep[D$\sim$180-200~km; e.g.,][]
  {Tedesco:2004um,Masiero:2011jc} %accounting for about half a percent of the 
  %mass of the asteroid belt \citep{DeMeo:2013hw,DeMeo:2014hk}
  that has long received particular attention from the community
  of asteroid spectroscopists, meteoricists, and dynamicists.
  Indeed, Hebe's spectral properties and close proximity to orbital resonances
  %the 3:1
  %mean-motion and the $\nu_6$ secular resonances 
  in the asteroid belt make it a possible
  main source of ordinary H~chondrites (i.e., $\sim$34\% of
  the meteorite falls, \citealt{Hutchison:2004uk, Farinella:1993fp,
   Migliorini:1997hp, Gaffey:1998ck, Bottke:2010vm}). 
   It was further proposed that Hebe could be the parent
   body of an ancient asteroid family \citep{Gaffey:2013vf}. The idea of H chondrites mainly 
   originating from Hebe, however, was recently 
   weakened by the discovery of a large number of asteroids
   (including several asteroid families) with similar spectral
   properties \citep[hence composition,][]{Vernazza:2014uqa}.  
   %The existence of a putative Hebe-derived family, on the other hand, 
   %still remains elusive.
   Here, we challenge this hypothesis by studying the  three-dimensional shape and topography 
   of Hebe derived from disk-resolved observations.
   We observed Hebe throughout its rotation in
   order to derive its shape, and to characterize the largest craters at its surface.
   When combined with previous adaptive-optics (AO) images and light curves (both from the
   literature and from recent optical observations by our team), 
   these new observations allow us to derive 
   a reliable shape model and an estimate of Hebe's density based on
   its astrometric mass (i.e., the mass derived from the study of planetary ephemeris and orbital deflections). 
   Finally, we analyse Hebe's topography by means of
   an elevation map and discuss the implications for the origin of H chondrites.

  % and the hypothesis of an ancient Hebe-derived family.

%  \indent Early adaptive-optics (AO) images of (6)~Hebe obtained at the
%  W.M. Keck II telescope have revealed non-convex features possibly
%  indicative of a violent collisional disruption in the asteroid's
%  history \citep{Hanus:2013jh}. However, 3-D modelling of its shape have
%  so far failed to determine a solution able to reproduce the non-convex
%  features of the contour, which was interpreted as the effect of the
%  shadow of some terrain relief creating apparent surface features on
%  the images \citep[observations were obtained at a relatively large phase
%  angle $\sim$30$\degree$;][]{Hanus:2013jh}. \\
%
 % \indent Here, we present new AO images of Hebe obtained with SPHERE
 % as part of the science verification of the instrument. Hebe was
 % observed close to its opposition date and throughout its rotation in
 % order to derive its shape, and to study the largest craters of its surface.
 % When combined to previous AO images, lightcurves (both from the
 % literature and from recent optical observations by our team), and
 % stellar occultations, these new observations allow us to derive 
 % a reliable shape model and an estimate of Hebe's density based on
 % its astrometric mass. Finally, we analyse Hebe's topography by mean of
 % an elevation map and discuss the implication for the origin of H chondrites
 % and an ancient asteroid family.
  
  %a large dataset of
  %thermal infrared measurements through thermophysical modelling 
  %to validate the size of our shape model, and to determine the
  %thermal properties of its surface.

\section{Observations and data pre-processing}

  \indent We observed (6)~Hebe close to its opposition date while it
  was orientated ``equator-on'' (from its spin solution derived
  below), that is, with an ideal viewing geometry
  exposing its whole surface as it rotated. Observations were acquired at four
  different epochs between December 8-12, 2014, such that the variation
  of the sub-Earth point longitude was 90$\pm$30$\degr$ between each epoch.  

  \indent Observations were performed with the recently commissioned
  second-generation SPHERE instrument, mounted at the European
  Southern Observatory (ESO) Very Large Telescope (VLT)
  \citep{Fusco:2006gm,Beuzit:2008gt}, during the science verification of  
  the instrument\footnote{Observations obtained under ESO programme ID
    60.A-9379 (P.I. C. Dumas)}. We used IRDIS broad-band classical
  imaging in Y (filter central wavelength=1.043\,$\mu$m, width=0.140\,$\mu$m) 
  in the pupil-tracking mode, where the pupil remains
  fixed while the field orientation varies during the observations, 
  to achieve the best point-spread function (PSF) stability.
  Each observational sequence
  consisted in a series of ten images with 2\,s exposure time during
  which Hebe was used as a natural guide star for AO
  corrections. Observations were performed under average seeing
  conditions (0.9-1.1\arcsec) and clear sky transparency, at an airmass of
  $\sim$1.1.

  \indent Sky backgrounds were acquired along our observations for
  data-reduction purposes. At the end of each sequence, we observed the
  nearby star HD~26086 under the exact same AO configuration as the
  asteroid to estimate the instrument PSF for deconvolution purposes.
  Finally, standard
  calibrations, which include 
  detector flat-fields and darks, were acquired in the morning as part of the
  instrument calibration plan.  

  \indent Data pre-processing of the IRDIS data made use of the preliminary
  release (v0.14.0-2) of the SPHERE data reduction and handling (DRH)
  software \citep{Pavlov:2008di}, as well as additional tools written
  in the Interactive Data Language (IDL), in order to perform
  background subtraction, flat-fielding and bad-pixel correction. The
  pre-processed images were then aligned one with respect to the others
  using the IDL ML\_SHIFTFINDER maximum likelihood function, 
  and averaged to maximise the signal to noise
  ratio of the asteroid. Finally, the optimal angular resolution of
  each image ($\lambda$/D=0.026", corresponding to a projected distance of 22\,km) 
  was restored with Mistral, a myopic deconvolution
  algorithm optimised for images with sharp boundaries
  \citep{Fusco:2002iz, Mugnier:2004kf}, using the stellar PSF acquired
  on the same night as our asteroid data.\\ 

\begin{table*}[!t]
\begin{center}
  \caption{
    Date, mid-observing time (UTC), 
    heliocentric distance ($\Delta$) and range to observer ($r$),
    phase angle ($\alpha$), apparent size ($\Theta$), longitude ($\lambda$) and latitude ($\beta$) 
    of the subsolar and subobserver points (SSP, SEP).
    PIs of these observations were
    $^1$J.-L. Margot,
    $^2$W. J. Merline,
    $^3$W. M. Keck engineering team,
    $^4$F. Marchis,
    $^5$B. Carry,
    and
    $^6$C. Dumas.
    \label{tab:obscondition}
  }
  \begin{tabular}{rcclrrrrrrrr}

    \hline\hline
     & Date & UTC & \multicolumn{1}{c}{Instrument} & \multicolumn{1}{c}{$\Delta$} & \multicolumn{1}{c}{$r$} & \multicolumn{1}{c}{$\alpha$} &
     \multicolumn{1}{c}{$\Theta$} &
     \multicolumn{1}{c}{SEP$_\lambda$} &
     \multicolumn{1}{c}{SEP$_\beta$} &
     \multicolumn{1}{c}{SSP$_\lambda$} &
     \multicolumn{1}{c}{SSP$_\beta$} \\
    &&&& \multicolumn{1}{c}{(AU)} & \multicolumn{1}{c}{(AU)} &
    \multicolumn{1}{c}{(\degr)} & \multicolumn{1}{c}{(\arcsec)} & 
    \multicolumn{1}{c}{(\degr)} &\multicolumn{1}{c}{(\degr)} &\multicolumn{1}{c}{(\degr)} &\multicolumn{1}{c}{(\degr)} \\
    \hline
  1&2002-05-07&14:08:54& Keck/NIRC2$^1$ &  2.52&  1.88& 20.5& 0.131&  66.1& -34.4&  53.3& -17.4 \\
  2&2002-05-08&13:55:01& Keck/NIRC2$^1$ &  2.52&  1.86& 20.4& 0.119& 329.8& -34.5& 317.2& -17.5 \\
  3&2002-09-27&06:29:51& Keck/NIRC2$^2$ &  2.21&  1.91& 27.0& 0.098& 162.5& -19.4& 187.2& -35.4 \\
  4&2007-12-15&14:15:39& Keck/NIRC2$^3$ &  2.47&  1.86& 20.8& 0.149&  14.2&  32.9& 356.2&  19.6 \\
  5&2007-12-15&14:30:31& Keck/NIRC2$^3$ &  2.47&  1.86& 20.8& 0.145&   1.9&  32.9& 343.9&  19.6 \\
  6&2007-12-15&14:44:49& Keck/NIRC2$^3$ &  2.47&  1.86& 20.8& 0.145& 350.1&  32.9& 332.1&  19.6 \\
  7&2007-12-15&15:00:54& Keck/NIRC2$^3$ &  2.47&  1.86& 20.8& 0.138& 336.8&  32.9& 318.9&  19.6 \\
  8&2007-12-15&15:27:39& Keck/NIRC2$^3$ &  2.47&  1.86& 20.8& 0.143& 314.8&  32.9& 296.8&  19.6 \\
  9&2007-12-15&16:26:58& Keck/NIRC2$^3$ &  2.47&  1.86& 20.8& 0.151& 265.9&  32.9& 247.9&  19.6 \\
 10&2009-06-07&10:52:24& Keck/NIRC2$^2$ &  2.81&  2.01& 15.1& 0.129&  43.1&   8.0&  57.9&   4.9 \\
 11&2010-06-28&13:08:00& Keck/NIRC2$^4$ &  2.06&  1.62& 28.9& 0.168& 258.8& -39.3& 221.2& -39.0 \\
 12&2010-08-26&12:47:10& Keck/NIRC2$^3$ &  1.98&  1.05& 16.1& 0.260&  48.5& -27.4&  30.6& -31.2 \\
 13&2010-08-26&13:04:26& Keck/NIRC2$^3$ &  1.98&  1.05& 16.1& 0.260&  34.3& -27.4&  16.3& -31.2 \\
 14&2010-08-26&13:59:47& Keck/NIRC2$^3$ &  1.98&  1.05& 16.1& 0.265& 348.6& -27.4& 330.7& -31.2 \\
 15&2010-08-26&14:38:00& Keck/NIRC2$^3$ &  1.98&  1.05& 16.1& 0.270& 317.1& -27.4& 299.2& -31.2 \\
 16&2010-11-29&07:10:28& Keck/NIRC2$^4$ &  1.94&  1.39& 28.9& 0.189& 160.9& -22.9& 191.5& -18.5 \\
 17&2010-12-13&01:18:16& VLT/NACO$^5$   &  1.94&  1.52& 30.0& 0.153&  28.9& -23.2&  59.6& -15.3 \\
 18&2010-12-13&02:40:24& VLT/NACO$^5$   &  1.94&  1.53& 30.0& 0.171& 321.1& -23.2& 351.8& -15.3 \\
 19&2010-12-14&00:41:59& VLT/NACO$^5$   &  1.94&  1.53& 30.0& 0.171& 311.5& -23.2& 342.2& -15.1 \\
 20&2010-12-14&01:38:22& VLT/NACO$^5$   &  1.94&  1.54& 30.0& 0.158& 265.0& -23.2& 295.7& -15.1 \\
 21&2010-12-14&02:14:10& VLT/NACO$^5$   &  1.94&  1.54& 30.0& 0.167& 235.5& -23.2& 266.2& -15.0 \\
 22&2014-12-08&00:53:28& VLT/SPHERE$^6$ &  2.03&  1.15& 17.0& 0.216& 208.7&   3.4& 225.7&   2.8 \\
 23&2014-12-09&01:04:54& VLT/SPHERE$^6$ &  2.03&  1.16& 17.2& 0.211&  91.6&   3.2& 108.9&   3.0 \\
 24&2014-12-10&01:59:38& VLT/SPHERE$^6$ &  2.03&  1.17& 17.5& 0.221& 298.8&   3.0& 316.4&   3.2 \\
 25&2014-12-12&04:14:08& VLT/SPHERE$^6$ &  2.04&  1.18& 18.1& 0.221& 332.6&   2.6& 350.7&   3.7 \\
  \hline
  \end{tabular}
\end{center}
\end{table*}

\FloatBarrier

\section{Additional data}

  \subsection{Disk-resolved images}

    \indent To reconstruct the
    3D shape of (6)~Hebe, we compiled available images
    obtained with the earlier-generation AO instruments 
    %NIRI \citep{Hodapp:2003ko} on the Gemini North telescope, 
    NIRC2 \citep{Wizinowich:2000ua, vanDam:2004jd} on the W. M. Keck II telescope 
    and NACO \citep{Lenzen:2003iu, Rousset:2003hh} on the ESO VLT. 
    Each of these images, as well as the corresponding
    calibration files and stellar PSF, were retrieved from the
    Canadian Astronomy Data
    Center\footnote{http://www.cadc-ccda.hia-iha.nrc-cnrc.gc.ca/} 
    \citep{Gwyn:2012gv} or
    directly from the observatory's database. Data processing and
    Mistral deconvolution of these
    images  were performed following the
    same method as for our SPHERE images. 
    Only a subset of the \numb{25} different epochs
    listed in Table~\ref{tab:obscondition} had been published
    \citep{Hanus:2013jh}.

  \subsection{Optical light curves}

    \indent We used \numb{38} light curves obtained in the years
    1953-1993 and available in the Database of Asteroid Models from
    Inversion Techniques
    \citep[DAMIT\footnote{http://http://astro.troja.mff.cuni.cz/projects/asteroids3D},][]{Durech:2010ey} that were used by 
    \citet{Torppa:2003ii} to derive
    the pole orientation and convex shape of (6)~Hebe from light curve
    inversion \citep{Kaasalainen:2001dx, Kaasalainen:2001di}.
    We also retrieved \numb{16} light curves observed by the amateurs
    F.~Kugel and J.~Caron, from the \textsl{Courbe de Rotation}
    group\footnote{http://obswww.unige.ch/{\textasciitilde}behrend/page\_cou.html}, and 
    84 light curves from the data archive of the SuperWASP survey
    \citep{Pollacco:2006gb} for the
    period 2006-2009. 
    This survey aims at finding and characterizing exoplanets by
    observation of their transit in front of their host star.
    Its large field of view and cadence provides a goldmine
    for asteroid light curves \citep{2017-ACM-Grice}.
    Finally, four light curves were acquired by our
    group during April 2016 with the 60\,cm TRAPPIST telescope
    \citep{Jehin:2011vb}.

  \subsection{Stellar occultations}

    \indent We retrieved the five stellar occultations listed by
    \citet{Dunham:2016wz} and publicly available on the Planetary Data
    System (PDS)\footnote{http://sbn.psi.edu/pds/resource/occ.html} for
    (6)~Hebe. We convert the disappearance and reappearance timings
    of the occulted stars into segments (called chords) on the plane
    of the sky, using the location of the observers on Earth and the
    apparent motion of Hebe following the recipes by \citet{1999-IMCCE-Berthier}.
    Of the five events, only two had more than one positive chord (that
    is a recorded blink event) and
    could be used to constrain the 3D shape (1977-03-05 -- also presented in \citealt{Taylor:1978jw} -- and
    2008-02-20).

  \subsection{Mid-infrared thermal measurements\label{sec:tm}}

    \indent Finally,  we compiled available mid-infrared thermal measurements to
    1) validate, independently of the AO images, the size of our 3D-shape model and,  
    2) derive the thermal properties of the surface of Hebe
    through thermophysical modeling of the infrared flux. 
    Specifically, we used a total of 103 thermal data points from
    IRAS \citep[12, 25, 60, 100\,$\mu$m,][]{Tedesco:2002fs},
    AKARI-IRC \citep[9, 18\,$\mu$m,][]{Usui:2011kh}, 
    ISO-ISOPHOT \citep[25\,$\mu$m,][]{Lagerros:1999ku}, and
    %Spitzer-MIPS (71, 156\,$\mu$m, \citealt{Gordon:2007ds, Stansberry:2007iqb}), 
    Herschel-PACS (70, 100, 160\,$\mu$m, M\"uller et al., in prep).
   % Herschel-SPIRE (250/350/500 micron) NOT USED
   % Planck (857 GHz) NOT USED
    %MSX (\citealt{Tedesco:2002gr}). 
    %WISE data were not included as highly saturated for Hebe. 
      
\section{3D shape, volume, and density\label{sec:3d}}

  \indent Recent algorithms such as KOALA \citep{Carry:2010bo, Carry:2012jn,
    Kaasalainen:2011tk} and ADAM \citep{Viikinkoski:2015jha} allow 
    simultaneous derivation of the spin, 3D shape, and size of an asteroid
  \citep[see, e.g.,][]{Merline:2013fs, Tanga:2015fx,
    Viikinkoski:2015jk, Hanus:2017ed}. 
  This combined multi-data approach has been validated by comparing
  the 3D shape model of (21) Lutetia by \citet{Carry:2010je} with the
  images returned by the ESA Rosetta mission during its fly-by of the
  asteroid \citep[see][]{Sierks:2011il,Carry:2012jn}.

  \indent Here, we reconstruct the spin and shape of
  (6)~Hebe with ADAM, which iteratively improves the solution by
  minimizing the residuals between the Fourier transformed images 
  and a projected polyhedral model. This method allows the use of AO images directly 
  without requiring the extraction of boundary contours. Boundary contours are therefore 
  used here only as a means to measure the pixel root mean square (RMS) residuals 
  between the location of the asteroid silhouette on the observed
  and modeled images. ADAM offers two different shape supports: subdivision surfaces 
  and octanoids based on spherical harmonics. Here, we use the subdivision surfaces 
  parametrisation which offers more local control 
  on the model than global representations (see \citealt{Viikinkoski:2015jha}).

\begin{figure}[h!]
\centering
\includegraphics[angle=0, width=0.7\linewidth, trim=0cm 0cm 0cm 0cm, clip]{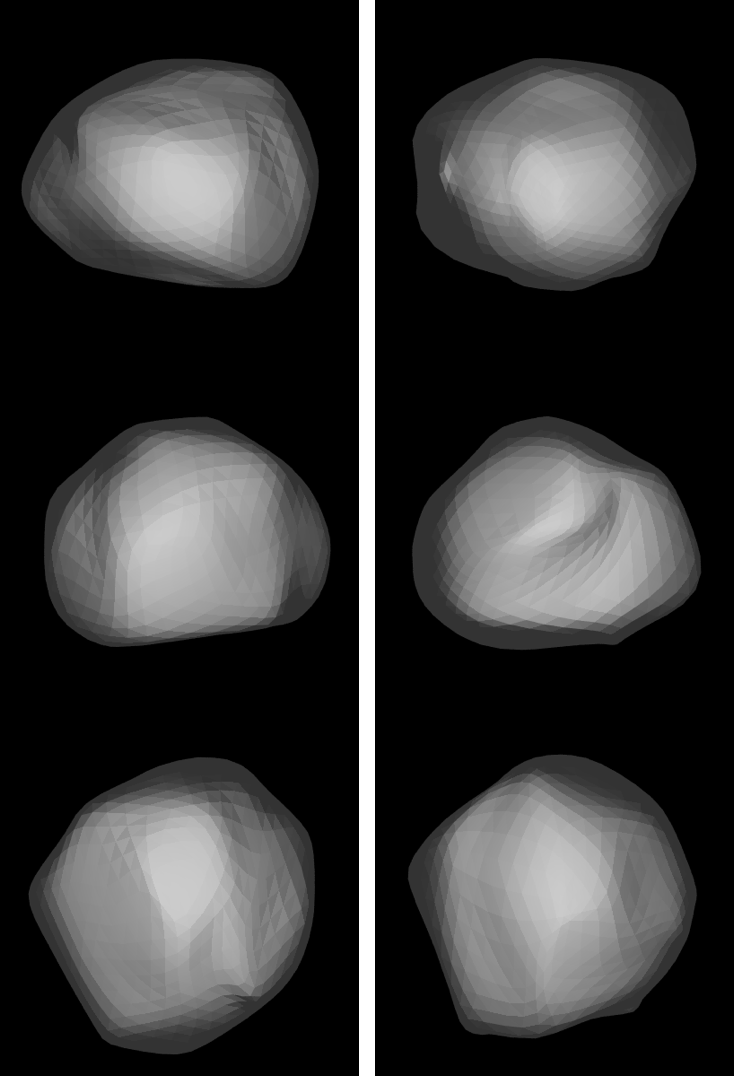}
 \caption[ ]{3D-shape model of (6)~Hebe reconstructed from light curves and all resolved images ({\it left}), and from light curves and SPHERE images only ({\it right}). Viewing directions are two equator-on views rotated by 90$\degr$ and a pole-on view.} 
\label{fig:model}
\end{figure}

\begin{figure}[h!]
\centering
\includegraphics[angle=0, width=\linewidth, trim=0cm 0cm 0cm 0cm, clip]{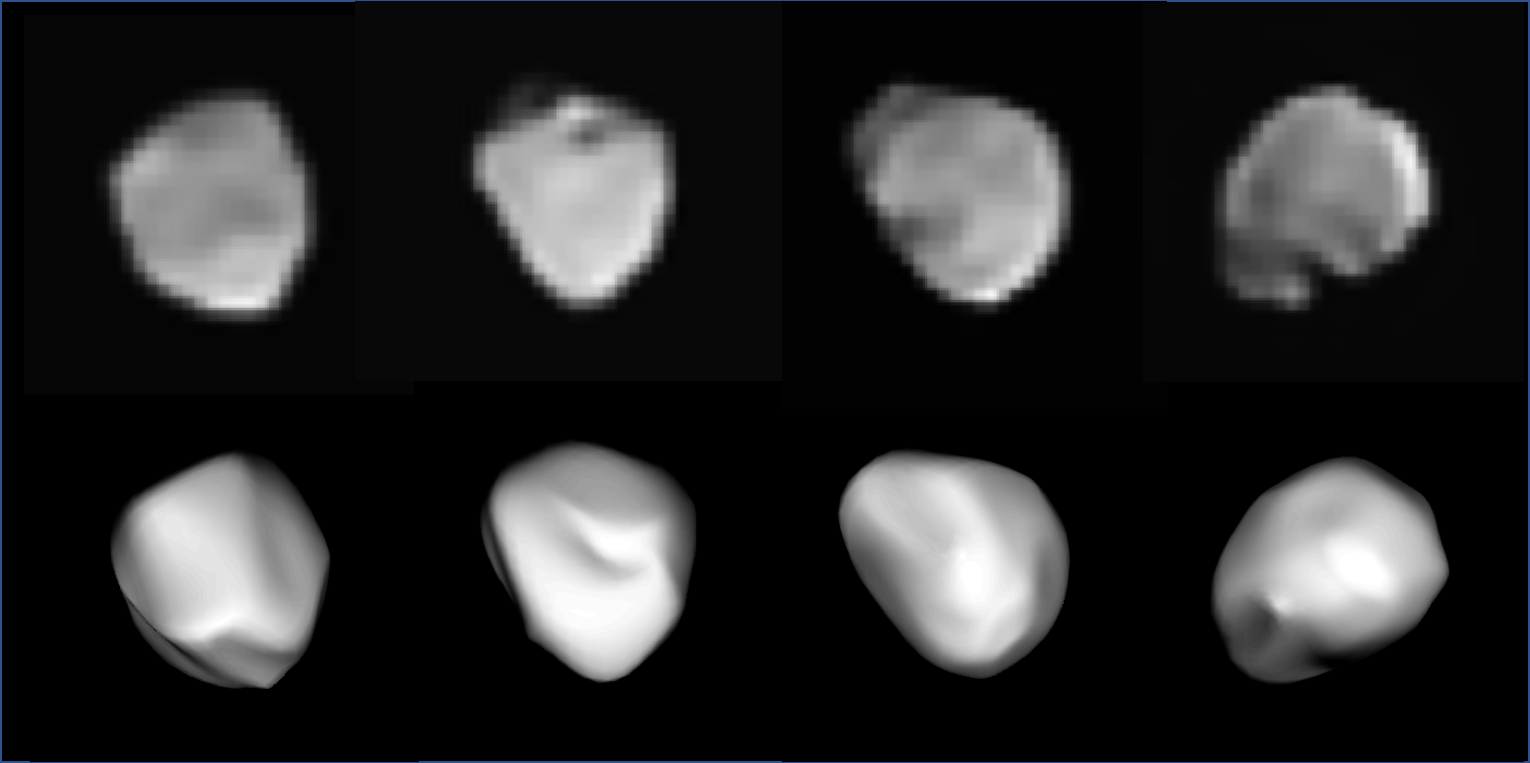}
 \caption[ ]{Deconvolved SPHERE images of Hebe obtained between 8 and 12 December 2014 ({\it top}) and corresponding projection of the model ({\it bottom}). Orientation of the four images with respect to the North is 15.2$\degree$, 12.8$\degree$, -5.3$\degree$ and -89.6$\degree$ , respectively.
 } 
\label{fig:proj}
\end{figure}

\begin{figure*}[!htb]
   \includegraphics[angle=0, width=\linewidth, trim=0cm 13.5cm 0cm 0cm, clip]{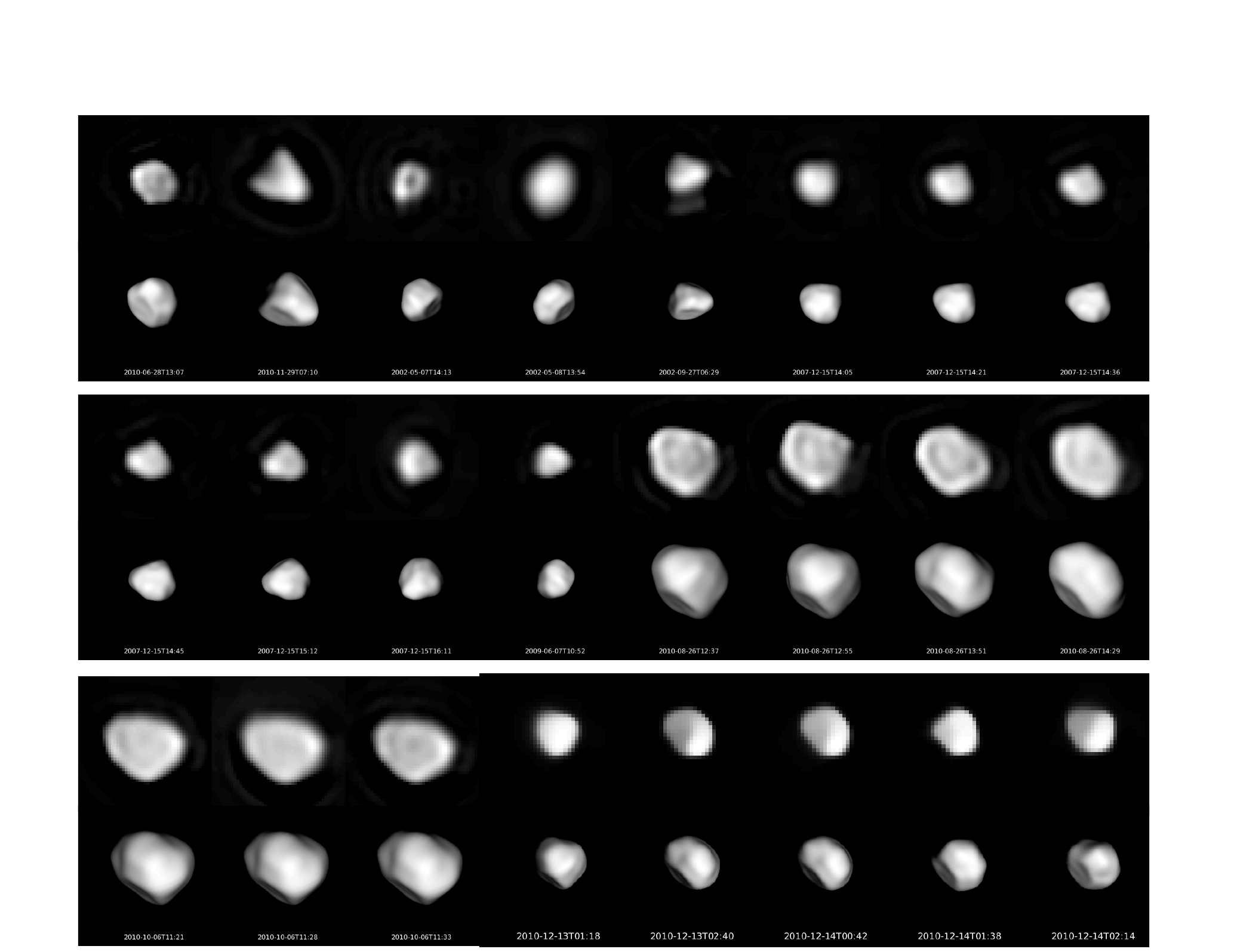}
    \caption[ ]{Previous AO images of Hebe obtained with Keck/NIRC2 and VLT/NACO ({\it top of the three rows}) and corresponding projection of the model ({\it bottom}). Each image is 0.8"$\times$0.8" in size.}
\label{fig:ao}
\end{figure*}

\begin{figure}[h!]
\centering
\includegraphics[angle=0, width=\linewidth, trim=0cm 0cm 0cm 0cm, clip]{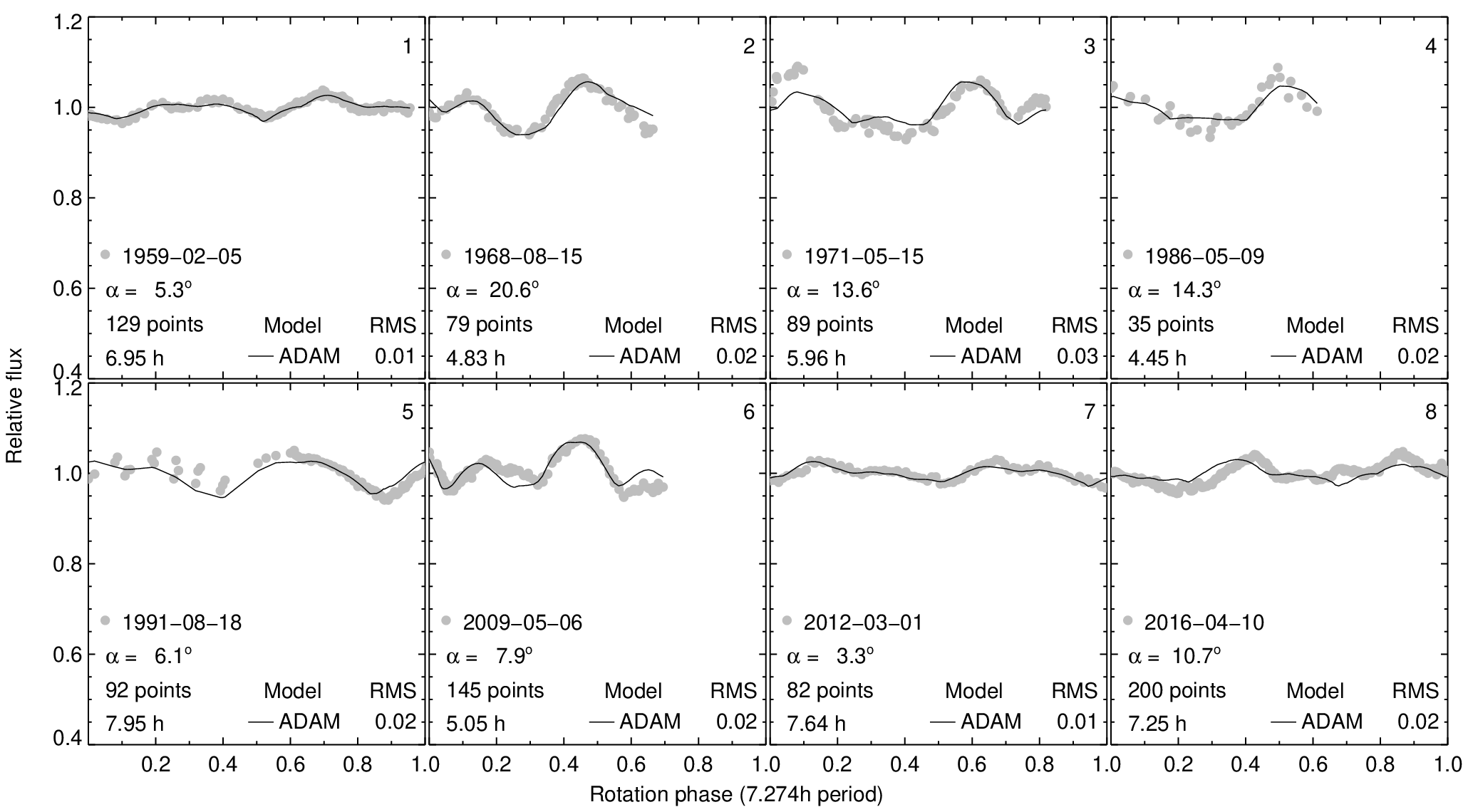}
 \caption[ ]{Comparison of the synthetic light curves (solid line) from the shape
   model with a selection of light curves (gray points). }
\label{fig:lc}
\end{figure}

\begin{figure}[h!]
\centering
\includegraphics[angle=0, width=\linewidth, trim=0cm 0cm 0cm 0cm, clip]{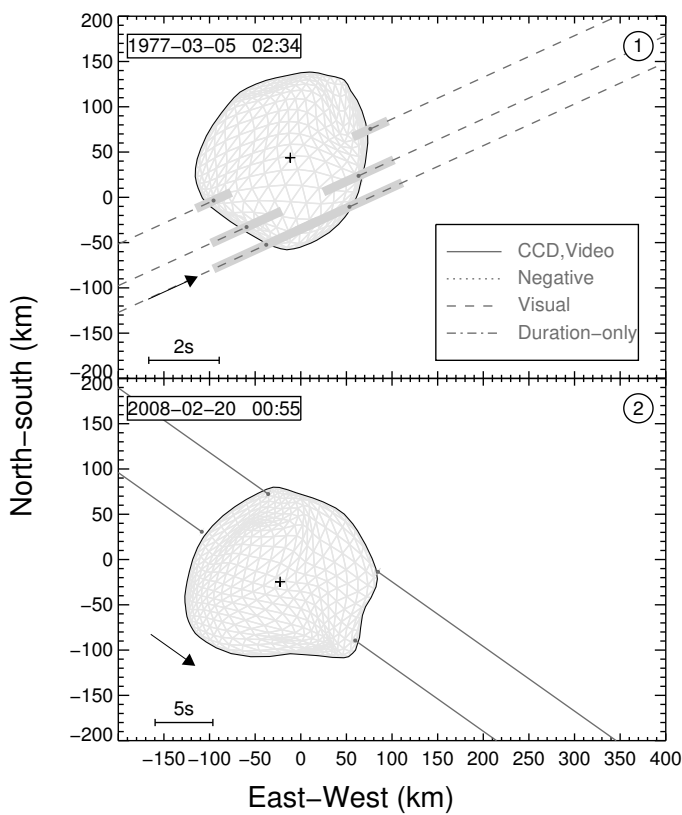}
 \caption[ ]{Comparison of the shape model with the chords from the occultation of 1977 and 2008. }
\label{fig:occ}
\end{figure}

  %We used \numb{58} lightcurves and \numb{23} AO images.
  Two different models depicted in Fig.~\ref{fig:model} were obtained; 
  the first one using the light curves combined to the full AO sample, 
  and the second one using the light curves and the SPHERE images only. 
  Comparison of the SPHERE-based model with our SPHERE images,
  earlier AO images, subsets of optical 
  light curves and stellar occultations are presented in
  Figs.~\ref{fig:proj}, \ref{fig:ao}, \ref{fig:lc}, and
  \ref{fig:occ}, respectively. 

  \indent The two models nicely fit all data, the 
  RMS residuals between the observations and the predictions by the model
  being only \numb{0.6} pixels for the location of the asteroid contours, 
  \numb{0.02} magnitude for the light curves, and \numb{5}\,km for the stellar occultation of 2008
  (the occultation of 1977 has very large uncertainties on its timings).
  The 3D shape models are close to an oblate spheroid, and have a
  volume-equivalent diameter of $196 \pm 6$\,km (all AO) and 
  $193 \pm 6$\,km (SPHERE-based; Table~\ref{tab:adam}). 
  Spin-vector coordinates ($\lambda$, $\beta$ in ECJ2000) are close
  to earlier estimates %\citep{Torppa:2003ii, Hanus:2013jh}.   
   based on light-curve inversion
  \citep[(339\degr,+45\degr),][]{Torppa:2003ii}
  and on a combination of light curves and AO images \citep[(345\degr,+42\degr),][]{Hanus:2013jh}. 
  
  The main difference between the two shape models comes from the presence of
  some surface features in the SPHERE-based model that are lacking
   in the model obtained using the full dataset of AO images. 
  This is due to the lower resolution of earlier AO images 
  that do not address some of the small-scale surface features revealed by the 
  SPHERE images. 
  
 %-----------------------------Figure Start-----------------------------
%-----------------------------Figure End ------------------------------

 %-----------------------------Figure Start-----------------------------
%-----------------------------Figure End ------------------------------

%-----------------------------Figure Start-----------------------------
%\begin{figure*}[h!]
%\centering
%\includegraphics[angle=0, width=\linewidth, trim=0cm 0cm 0cm 0cm, clip]{art-ao-001.eps}
% \caption[]{Comparison of the shape model (black contours, and gray
%   triangles describing the surface) with a selection of
%   profiles from AO images (gray shaded in the background).
%   The main feature revealed from contour \#8 is 
%   more likely due to a shadowing effect rather than a true contour variation.}
%\label{fig:ao}
%\end{figure*}
%-----------------------------Figure End ------------------------------

%-----------------------------Figure Start-----------------------------
%-----------------------------Figure End ------------------------------

%-----------------------------Figure Start-----------------------------
%-----------------------------Figure End ------------------------------

%-----------------------------Figure Start-----------------------------
%-----------------------------Figure End ------------------------------

%%%%%%%%%%%%%%%%%%%%%%%%%%%%%%%%%%%%%%%%%%%%%%%%%%%%%%%%%%%%%%%%%%%%%%%%%%%%%%%%%%%%%%%%%%%%%%%%%%%%%%
%\begin{table}[ht]
%\begin{center}
%  \caption{Period, spin (ECJ2000 longitude $\lambda$, latitude $\beta$ and initial Julian date T$_0$) and dimensions
%  (volume-equivalent diameter $D$, volume $V$, and tri-axial ellipsoid diameters $a$, $b$, $c$ along 
%  principal axes of inertia) of Hebe derived with KOALA.
%  \label{tab:koala}
%  }
%  \begin{tabular}{llll}
%    \hline\hline
%    Parameter & Value & Unc. & Unit \\
%    \hline
%    Period    & 7.274465 & 5.10$^{-5}$ & hour \\
%    $\lambda$ & 345.9 & 180.0 & deg. \\
%    $\beta$   & 51.7 & 22.0 & deg. \\
%    T$_0$     & 2434569.00000000 & & \\
%    \hline
%    D       & 196.48 & 11.31 & km \\
%    V       & 4.10$^{6}$ & 2.10$^{5}$ & km$^3$ \\
%    a       & 212.9 & 11.3 & km\\
%    b       & 203.3 & 11.3 & km\\
%    c       & 178.0 & 11.3 & km\\
%    a/b     & 1.05 & 0.08 & \\
%    b/c     & 1.14 & 0.08 & \\
%    \hline
%  \end{tabular}
%\end{center}
%\end{table}
%%%%%%%%%%%%%%%%%%%%%%%%%%%%%%%%%%%%%%%%%%%%%%%%%%%%%%%%%%%%%%%%%%%%%%%%%%%%%%%%%%%%%%%%%%%%%%%%%%%%%%

%%%%%%%%%%%%%%%%%%%%%%%%%%%%%%%%%%%%%%%%%%%%%%%%%%%%%%%%%%%%%%%%%%%%%%%%%%%%%%%%%%%%%%%%%%%%%%%%%%%%%%
\begin{table}[ht]
\begin{center}
  \caption{Period, spin (ECJ2000 longitude $\lambda$, latitude $\beta$ and initial Julian date T$_0$), and dimensions
  (volume-equivalent diameter $D$, volume $V$, and tri-axial ellipsoid diameters $a$, $b$, $c$ along 
  principal axes of inertia) of Hebe derived with ADAM.
  \label{tab:adam}
  }
  \begin{tabular}{llllll}
    \hline\hline
    Parameter & Value & Value & Unc. & Unit \\
     & \multicolumn{1}{l}{(all AO)} & \multicolumn{1}{l}{(SPHERE-only)} & & & \\[1mm]
    \hline
    Period    & 7.274467 & 7.274465 & 5.10$^{-5}$ & hour \\
    $\lambda$ & 341.7 & 343.2 & 3 & deg. \\
    $\beta$   & +49.9 & +46.8 & 4 & deg. \\
    T$_0$     & 2434569.00 & 2434569.00 & & \\
    \hline
    D       & 196 & 193 & 6 & km \\
    V       & 3.95 $\cdot 10^{6}$ & 3.75 $\cdot 10^{6}$ & 1.2 $\cdot 10^{5}$ & km$^3$ \\
    a       & 218.4  & 213.4 & 6.0 & km \\
    b       & 206.2   & 200.2 & 6.0 & km \\
    c       & 172.1  & 172.6 & 6.0 & km \\
    a/b     & 1.06   & 1.07 & 0.04 & \\
    b/c     & 1.20   & 1.16 & 0.05 & \\
    \hline
  \end{tabular}
\end{center}
\end{table}
%%%%%%%%%%%%%%%%%%%%%%%%%%%%%%%%%%%%%%%%%%%%%%%%%%%%%%%%%%%%%%%%%%%%%%%%%%%%%%%%%%%%%%%%%%%%%%%%%%%%%%

  \indent There are 12 diameter estimates for Hebe in the literature
  (Table~\ref{table:diamestimates},
  Figure~\ref{fig:hebe-diam}). Rejecting values that do not fall
  within one standard deviation of the average value of the full
  dataset gives an average equivalent-volume sphere diameter of 
  $191.5 \pm 8.3$\,km, in very good agreement with the values 
  of \numb{$193 \pm 6$}\,km and \numb{$196 \pm 6$}\,km derived here 
  (also supported by the
  thermophysical analysis presented in the following section).
  In the following, we use the value of the diameter obtained from our
  SPHERE-based model, which is more precise 
  due to the higher angular resolution
  of the SPHERE images with respect to the NIRC2 and NACO images.
  %which is more precise and based on the largest data set of disk-resolved data. 
  A main advantage of using a diameter obtained from a full 3D
  shape modeling resides in the uncertainty on the derived volume $V$,
  which is close to $\delta V / V \approx \delta D / D$, as opposed to
  a $\delta V / V \approx 3 \delta D / D$ in the spherical assumption
  used in most aforementioned estimates
  (see \citealt{Kaasalainen:2012ku} for details).

%densite: 1.31e19/(4./3.*num.pi*(194000./2.)**3.)
%incertitude densite: num.sqrt((0.24/1.31)**2.+(11.71/195.67)**2.)*3.33
%porosite: 100.*(1.-3.33/3.42)
%incertitude porosite: num.sqrt(((0.18/3.42)**2.+(0.83/3.33)**2.))*2.
  \indent Combining this diameter with an average mass of 
  $1.31\pm0.24\times10^{19}$~kg computed from 16 estimates
  gathered from the literature (Table~\ref{table:massestimates}, Figure~\ref{fig:hebe-mass}), 
  provides a bulk density of $3.48\pm0.64$~g.cm$^{-3}$, 
  in perfect agreement with
  the average grain density of ordinary H~chondrites ($
  3.42\pm0.18 $~g.cm$^{-3}$; \citealt{Consolmagno:2008cl}). 
  The derived density therefore suggests a null internal porosity, 
  consistent with an intact internal structure. 
  Hebe hence appears to reside in the volumetric and
  structural transitional region between the compact and
  gravity-shaped dwarf planets, and the medium-sized asteroids
  ($\sim$10-100 km in diameter) with fractured interior
  \citep{Carry:2012cw,2015-AsteroidsIV-Scheeres}.
  However, due to the current large mass uncertainty that
  dominates the uncertainty of the bulk density,
  the possibility of higher internal porosity
  cannot be ruled out.
  We expect the Gaia mission to trigger higher-precision mass estimates 
  in the near future \citep{Mignard:2007er, Mouret:2007eg} that will help refine the density 
  measurement of Hebe.

\section{Thermal parameters and regolith grain size\label{sec:neatm}}

  \indent A thermophysical model (TPM; \citealt{Mueller:1998to, Muller:1999vg}) 
  was also used to provide an independent size measurement for Hebe 
  and to derive its thermal surface properties. 
  The TPM uses as input our 3D shape model with unscaled diameter. 
  The procedure is described in detail in Appendix\,\ref{sec:tpm}.
  
  Using absolute magnitude H=5.71 and magnitude slope G=0.27 
  from the Asteroid Photometric Catalogue \citep{Lagerkvist:2011vg}, the TPM
  provides a solution for diameter and albedo of ($D$, p$_{\rm v}$)=(198$\substack{+4 \\ -2}$\,km, 0.24$\pm$0.01), 
  in good agreement with the size of our 3D-shape model and previous albedo measurements from 
  IRAS (p$_{\rm v}$=0.27$\pm$0.01; \citealt{Tedesco:2002fs}), 
  WISE (p$_{\rm v}$=0.24$\pm$0.04; \citealt{Masiero:2014gu}) and 
  AKARI (p$_{\rm v}$=0.24 0.01; \citealt{Usui:2011kh}). 
  Best-fitting solutions are found for significant surface
  roughness (in agreement with \citealt{Lagerros:1999ku}), 
  and thermal inertia $\Gamma$ values ranging from 15 to
  90~J\,m$^{-2}$\,s$^{-0.5}$K\,$^{-1}$, with a preference for  
  $\Gamma \approx$\,50~J\,m$^{-2}$\,s$^{-0.5}$K\,$^{-1}$.  
  Interestingly, we note that the best-fitting solution for $\Gamma$
  drops from $\sim$60~J\,m$^{-2}$\,s$^{-0.5}$K\,$^{-1}$ 
  when only considering thermal measurements acquired at r$<$2.1~AU, 
  to $\sim$40~J\,m$^{-2}$\,s$^{-0.5}$K\,$^{-1}$ for data 
  taken at r$>$2.6 AU. While this might be indicative of changing thermal inertia with temperature, 
  this result should be taken with extreme caution, as the TPM probably overfits the data
  due to the large error bars on the thermal measurements (see Appendix\,\ref{sec:tpm}).

  %Remarkably, we note that the best-fitting solution for $\Gamma$
  %drops from $\sim$60~J\,m$^{-2}$\,s$^{-0.5}$K\,$^{-1}$ 
  %when only considering thermal measurements acquired at r$<$2.1~AU, 
  %to $\sim$40~J\,m$^{-2}$\,s$^{-0.5}$K\,$^{-1}$ for data 
  %taken at r$>$2.6 AU.
  %This is the first time such dependency of the thermal
  %inertia over heliocentric distance is ever observed for a solar system object. 
  %This behaviour is in perfect agreement with the temperature-dependancy 
  %of thermal conductivity and specific heat. Indeed, assuming a fine-grained lunar-like
  %regolith with $\Gamma$=60~J\,m$^{-2}$\,s$^{-0.5}$K\,$^{-1}$ at temperature T=230~K,
  %we find based on \citet{Keihm:1984hn} that T and $\Gamma$ should respectively 
  %drop to about 180~K and 45 J\,m$^{-2}$\,s$^{-0.5}$K\,$^{-1}$ at r$>$2.6 AU,
  %in very good agreement with our observations.
  
  From the thermal inertia value derived here, one can further derive the grain size of the 
  surface regolith of Hebe \citep{Gundlach:2013kn}. Assuming values of heat capacity and 
  material density typical of H5 ordinary chondrites \citep{Opeil:2010hq} and estimated
  surface temperature of 230\,K and 180\,K at 1.94 and 2.87\,AU respectively, we find that
  the typical grain size of Hebe is about 0.2--0.3\,mm (see Annexe\,\ref{sec:tpm} for more details).

\section{Topography}

\begin{figure*}[h!]
\centering
\includegraphics[angle=0, width=0.6\linewidth, trim=0cm 0cm 0cm 0cm, clip]{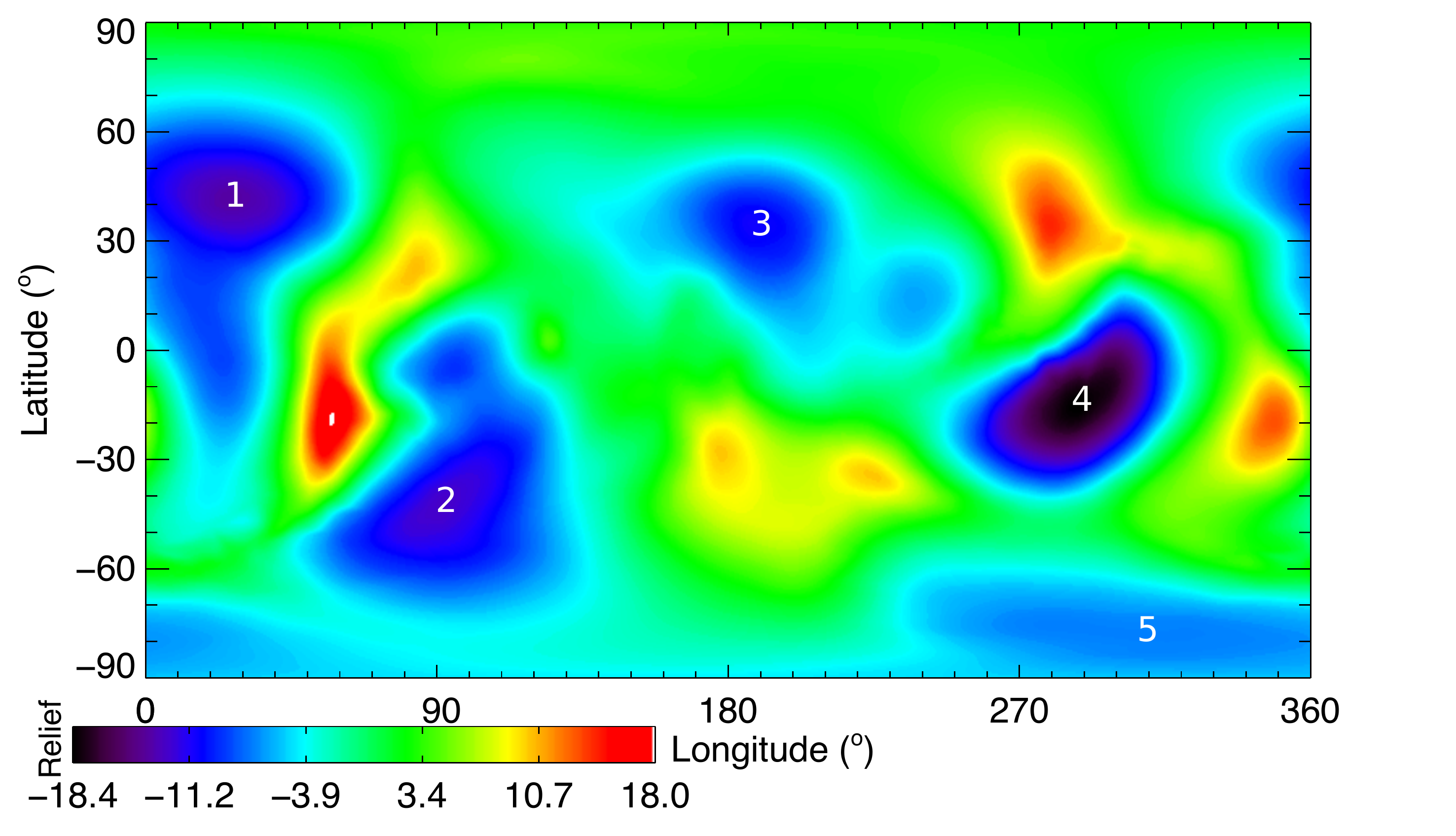}
 \caption[ ]{Elevation map (in km) of (6)~Hebe, with respect to a volume-equivalent ellipsoid best
   fitting our 3D-shape model. The five major depressions are identified by numbers.
 } 
\label{fig:topo}
\end{figure*}

  \indent Hebe's topography was investigated by generating an elevation
  map of its surface with respect to a volume-equivalent ellipsoid
  best-fitting our 3D-shape model, following the method by
  \citet{Thomas:1999jj}.
  This map shown in Figure\,\ref{fig:topo}
  allows the identification of several low-topographic and concave
  regions possibly created by impacts
  (the two shape models depicted in Fig.~\ref{fig:model} produce slightly
  different but consistent topographic maps). Specifically, five large 
  depressions (numbered 1 to 5 on the elevation map) are found at the surface of Hebe, at 
  (29$\degree$, 43$\degree$), (93$\degree$, -42$\degree$),
  (190$\degree$, 35$\degree$), (289$\degree$, -13$\degree$), and near
  the south pole. Estimated dimensions (diameter D and maximum depth below the average surface d) are 
  D$_1$=92--105\,km, d$_1$=13\,km; D$_2$=85--117\,km, d$_2$=12\,km; D$_3$=68--83\,km, d$_3$=11\,km;
  D$_4$=75--127\,km, d$_4$=18\,km; and D$_5$=42--52\,km, d$_5$=7\,km,  respectively.

  %All depressions are 90 to 130\,km in mean diameter and reach a
  %depth of $\sim$8 to 17\,km. 

  %Considering that relaxation does not
  %significantly reshape craters at the surface of rocky
  %(volatile-poor) asteroids, our elevation map first reveals that fragments
  %as large as $\sim$15\,km in diameter are unlikely to have originated
  %from Hebe. 
  % This possibility was raised by \citet{Gaffey:2013vf} who
  %proposed that (695)~Bella, (1166)~Sakuntala, and (1607)~Mavis, which
  %are respectively $41\pm1$, $26\pm1$, and $13\pm1$\,km in diameter
  %\citep{Masiero:2014gu}, could be members the putative Hebe family. 
  %If true, then we expect those bodies to be gravitationally bound 
  %conglomerates of smaller pieces. 

%An origin of these bodies from Hebe would require them to be gravitationally bound conglomerates of smaller fragments.

  \indent Assuming that the volume of a crater relates approximately to the
  volume of ejecta produced by the impact -- which is most likely very optimistic 
  because 1) a significant fraction of impact crater volume comes from 
  compression \citep{Melosh:1989uq} and, 2) at least a fraction of the ejecta must 
  have re-accumulated on the surface of the body (e.g., \citealt{Marchi:2015js}),
  one can further estimate the
  volume of a hypothetical family derived from an impact on Hebe. 
  The largest depression on Hebe roughly accounts for a volume of 
  $10^5 $~km$^3$, 
  corresponding to a body with an
  equivalent diameter of $\sim$58\,km. 
    
  For comparison, the five known S-type
  families spectrally analogous to Hebe (therefore to H chondrites; \citealt{Vernazza:2014uqa}) 
  and located close to the main-belt 3:1 and 5:2 mean-motion resonances,
  namely Agnia (located at semi-major axis $a$=2.78~AU and eccentricity
  $e$=0.09), Koronis ($a$=2.87~AU, $e$=0.05), Maria ($a$=2.55~AU,
  $e$=0.06), Massalia ($a$=2.41~AU, $e$=0.14) and Merxia ($a$=2.75~AU,
  $e$=0.13) encompass a total volume of respectively $ \sim 2.4 \times
  10^4 $~km$^3$, $ 5.6 \times 10^5 $~km$^3$, $ 3.6 \times 10^5 $~km$^3$,
  $ 5.7 \times 10^4 $~km$^3$ and $ 1.8 \times 10^4 $~km$^3$ when the
  larger member of each family is removed. Family membership was
  determined using \citet{Nesvorny:2015tp}'s Hierarchical Clustering
  Method (HCM)-based classification
  (http://sbn.psi.edu/pds/resource/nesvornyfam.html) and rejecting
  possible interlopers that do not fit the "V-shape" criterion as
  defined in \citet{Nesvorny:2015gu}. The diameter of each asteroid
  identified as a family member was retrieved from the WISE/NEOWISE
  database \citep{Masiero:2011jc} when available, or estimated from
  its absolute H~magnitude otherwise, assuming an albedo equal to that
  of the largest member of its family (respectively 0.152, 0.213,
  0.282, 0.241 and 0.213 for (847)~Agnia, (158)~Koronis, (170)~Maria,
  (20)~Massalia and (808)~Merxia; \href{https://mp3c.oca.eu}{https://mp3c.oca.eu}).
  We note that these values should be considered as lower limits as those families certainly
  include smaller members beyond the detection limit.

  \indent We therefore find that the volume of material 
  corresponding to the largest depression
  on Hebe is of the order of some H-chondrite-like S-type 
   families, and $\sim$4-6 times smaller 
  than the largest ones. 
 Therefore, although we cannot firmly exclude Hebe as the main (or unique) source of H chondrites, it appears that such a hypothesis is 
  not the most likely one.
  This is further strengthened by the following two arguments.
  First, it seems improbable that the volume excavated from Hebe's largest depression, 
  which we find to be roughly 10 to 30 times smaller than the volume of the Rheasilvia 
  basin on Vesta \citep{Schenk:2012in}, would contribute to $\sim$34\% of the meteorite
  falls, when HED meteorites only represent $\sim$6\% of the falls \citep{Hutchison:2004uk}. 
  We note, however, that the low number of HED meteorites may also relate to the relatively old age \citep{Schenk:2012in}
  of the Vesta family \citep{Heck:2017jg}.
  Second, the current lack of observational evidence for 
  a Hebe-derived family indicates that such a family, 
  if it ever existed, must be very ancient and dispersed. Yet, there is growing evidence from laboratory experiments that the current 
  meteorite flux must be dominated by fragments from recent asteroid breakups \citep{Heck:2017jg}. In the case of H chondrites, this is well supported 
  by their cosmic ray exposure ages  \citep{Marti:1992dv, Eugster:2006vg}. It therefore appears that a recent - yet to be identified - 
  collision suffered by another H-chondrite-like asteroid is the most
  likely source of the vast majority of H chondrites.

%-----------------------------Figure Start-----------------------------
%-----------------------------Figure End ------------------------------

\section{Conclusion and outlook}

  \indent We have reconstructed the spin and tridimensional shape of
  (6)~Hebe from combined AO images and optical light curves, and 
  checked the consistency of the derived model against available
  stellar occultations and thermal measurements.
  Whereas the irregular shape of Hebe suggests it was moulded by impacts,
  its density appears indicative of a compact interior. 
  Hebe thus seems to reside in the structural regime
  in transition between round-shaped dwarf planets shaped by gravity,
  and medium-sized asteroids with fractured interiors (i.e.,
  significant fractions of macro-porosity; \citealt{Carry:2012cw}).  
  This however needs to be confirmed by future mass measurements 
  (e.g., from Gaia high-precision astrometric measurements) 
  that will help improve the current mass uncertainty that
  dominates the uncertainty on density.

  %We further validated the size of our shape model by performing a
  %thermophysical modeling of available mid- to far-infrared
  %measurements, which also allowed us to derive the thermal properties
  %of the surface of Hebe. Remarkably, we report the first observation
  %of a correlation between thermal inertia and heliocentric distance
  %for a solar system object. 

  The high angular resolution of SPHERE further allowed us to
  identify several concave regions at the surface of Hebe possibly
  indicative of impact craters. %An analysis of their size and depth
  %reveals that these regions are unlikely to be the site of origin of
  %fragments as large as $\sim$15~km in
  %diameter. 
  We find the volume of the largest depression
  to be roughly five times smaller than
  the volume of the largest S-type H-chondrite-like families located close 
  to orbital resonances in the asteroid belt.
  Furthermore, this volume is  more than an order of magnitude smaller than 
  the volume of the Rheasilvia basin on Vesta \citep{Schenk:2012in}
  from which HED meteorites ($\sim$6\% of the falls) originate.
  Our results therefore imply that (6) Hebe is not the most likely source 
  of ordinary H chondrites ($\sim$34\% of the falls).

  Finally, this work has demonstrated the potential of SPHERE
  to bring important constraints on the origin and collisional history 
  of the main asteroid belt. 
  Over the next two years, our team will collect -- via a large
  program on VLT/SPHERE (run ID: 199.C-0074, PI: Pierre Vernazza) -- 
  similar volume, shape, and topographic measurements for 
  a significant number ($\sim$40) of D$\geq$100 km asteroids sampling the
  four major compositional classes (S, Ch/Cgh, B/C and P/D).

\begin{acknowledgements}

  Based on observations made with ESO Telescopes at the La Silla Paranal
  Observatory under programme ID 60.A-9379.
  The asteroid diameters and albedos based on NEOWISE observations
  were obtained from the Planetary Data System (PDS). 

  Some of the data presented herein were obtained at the W.M. Keck
  Observatory, which is operated as a scientific partnership among the
  California Institute of Technology, the University of California and
  the National Aeronautics and Space Administration. The Observatory
  was made possible by the generous financial support of the W.M. Keck
  Foundation.  

  This research has made use of the Keck Observatory Archive (KOA),
  which is operated by the W. M. Keck Observatory and the NASA
  Exoplanet Science Institute (NExScI), under contract with the
  National Aeronautics and Space Administration.  

  The authors wish to recognize and acknowledge the very significant
  cultural role and reverence that the summit of Mauna Kea has always
  had within the indigenous Hawaiian community.  We are most fortunate
  to have the opportunity to conduct observations from this mountain.  

  This research used the MP$^{3}$C service developed,
  maintained, and hosted at the Lagrange laboratory, Observatoire de la
  C{\^o}te d'Azur \citep{2017-ACM-Delbo}.
  
  Photometry of 6 Hebe was identified and extracted from WASP data 
  with the help of Neil Parley (Open University, now IEA Reading).
  The WASP project is currently funded and operated by Warwick University and Keele University, and was originally set up by Queen's University Belfast, the Universities of Keele, St. Andrews and Leicester, the Open University, the Isaac Newton Group, the Instituto de Astrofisica de Canarias, the South African Astronomical Observatory and by STFC.
  
  The WASP project is currently funded and operated by Warwick
  University and Keele University, and was originally set up by Queen's
  University Belfast, the Universities of Keele, St. Andrews and
  Leicester, the Open University, the Isaac Newton Group, the Instituto
  de Astrofisica de Canarias, the South African Astronomical Observatory
  and by STFC. 
  
  TRAPPIST-South is a project funded by the Belgian Funds (National) de la Recherche Scientifique (F.R.S.-FNRS) under grant FRFC 2.5.594.09.F, with the participation of the Swiss National Science Foundation (FNS/SNSF).
E.J. and M.G. are F.R.S.-FNRS research associates.

Based on observations with ISO, an ESA project with instruments
   funded by ESA Member States and with the participation of ISAS
   and NASA.
   
Herschel is an ESA space observatory with science instruments
   provided by European-led Principal Investigator consortia and
   with important participation from NASA.   
   
   Herschel fluxes of Hebe where extracted by Csaba Kiss (Konkoly Observatory, Research Centre for Astronomy
and Earth Sciences, Hungarian Academy of Sciences, H-1121 Budapest,
Konkoly Thege Mikl{\'o}s {\'u}t 15-17, Hungary).

  TM received funding from
  the European Union's Horizon 2020 Research and Innovation Programme,
  under Grant Agreement no 687378. 

\end{acknowledgements}

\bibliography{references}

\appendix

\section{Diameter and mass estimates from the literature}

\begin{figure*}[h!]
\centering
\includegraphics[angle=0, width=0.75\linewidth, trim=0cm 0cm 0cm 0cm, clip]{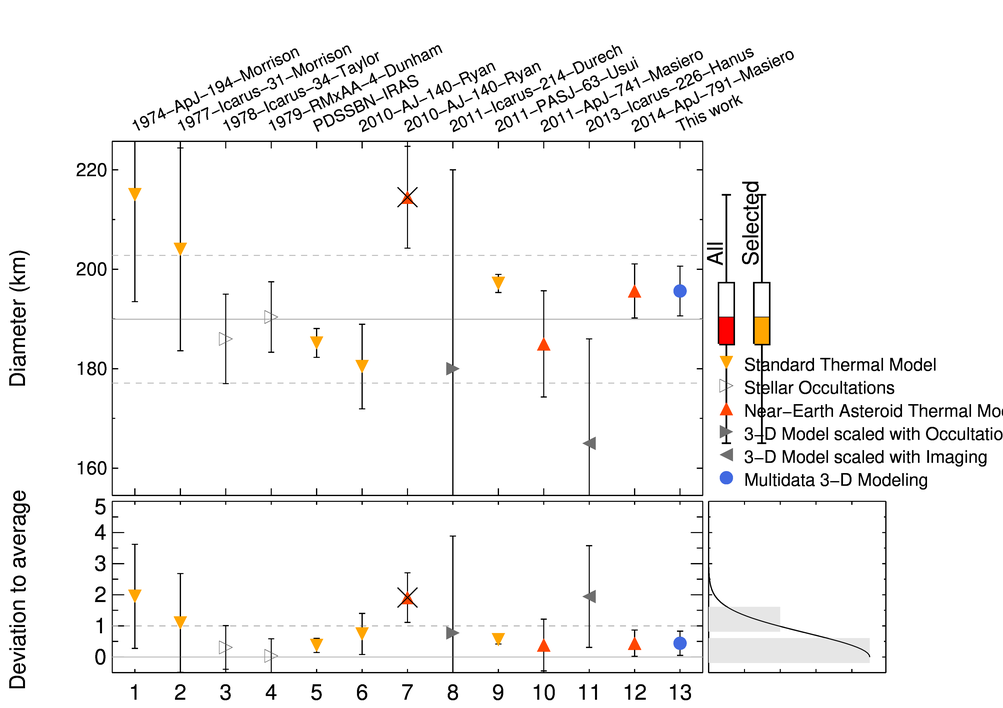}
 \caption[ ]{Diameter estimates of (6)~Hebe gathered from the literature.}
\label{fig:hebe-diam}
\end{figure*}

\begin{figure*}[h!]
\centering
\includegraphics[angle=0, width=0.75\linewidth, trim=0cm 0cm 0cm 0cm, clip]{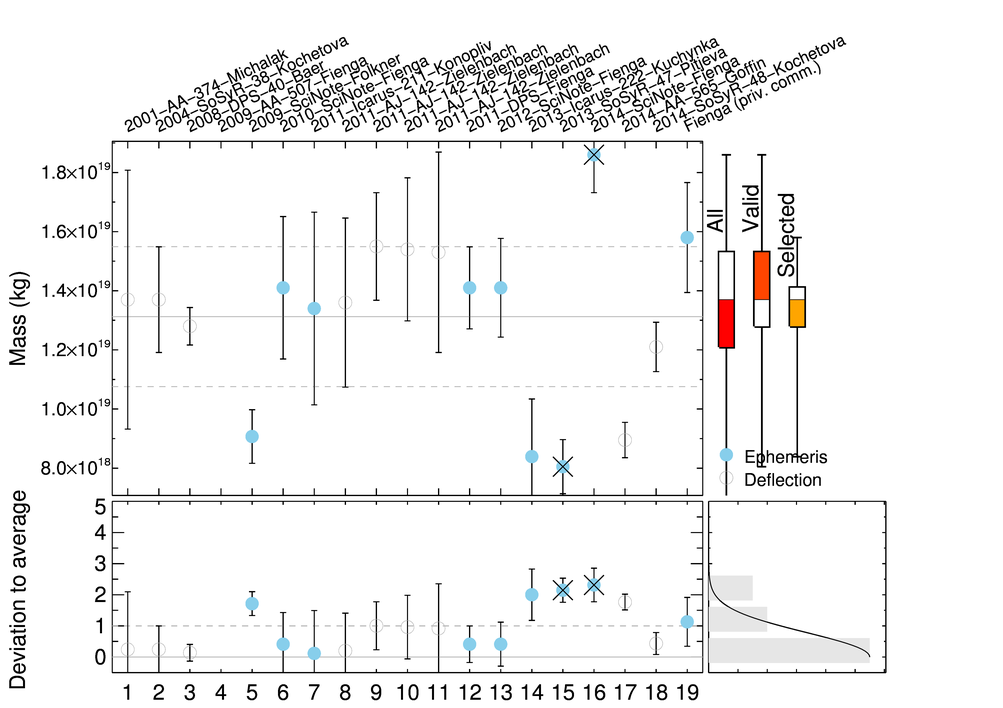}

 \caption[ ]{Mass estimates of (6)~Hebe gathered from the literature.}
\label{fig:hebe-mass}
\end{figure*}

Diameter and mass estimates of (6)~Hebe from the literature are
presented here in Table\,\ref{table:diamestimates} and Figure\,\ref{fig:hebe-diam} (diameter)
and Table\,\ref{table:massestimates} and Figure\,\ref{fig:hebe-mass} (mass).
Average values were determined following the method by \citet{Carry:2012cw}, 
which consists in rejecting all the estimates that do not
fall within one standard deviation of the average value, then by
recomputing the average without these values.

%-----------------------------Table Start-----------------------------
\begin{table*}[tbh!]
\footnotesize
\begin{center}
  \caption[]{\em{\small \rm Volume-equivalent diameter estimates of
      (6)~Hebe gathered from the literature. STM: Standard Thermal
      Model, NEATM: Near-Earth Asteroid Thermal Model, LC: light curve,
      Occ: stellar occultation, AO: adaptative optics imaging, LC+Occ:
      light curve-based 3-D model scaled using an occultation, LC+AO:
      light curve-based 3D model scaled using adaptative optics
      images.\\}} 
\label{table:diamestimates}
\begin{tabular}{rr@{\,$\pm$\,}rcl}
\hline \noalign {\smallskip}
  & \multicolumn{2}{c}{Diameter ($D$, km)} & Method & Reference  \\
\hline \noalign {\smallskip}
  1 & 215.00 & 21.50 & STM   & \citet{Morrison:1974en} \\
  2 & 201.00 & 20.10 & STM   &  \citet{Morrison:1977dc} \\
  3 & 186.00 & 9.00 & Occ   & \citet{Taylor:1978jw} \\
  4 & 190.40 & 7.10 & Occ   & \citet{Dunham:1979uh} \\
  5 & 185.18 & 2.90  & STM   & \citet{Tedesco:2004um} \\
  6 & 180.42 & 8.50  & STM   & \citet{Ryan:2010bu} \\
  7 & 214.49 & 10.25 & NEATM & \citet{Ryan:2010bu} \\
  8 & 180.00 & 40.00 & LC+Occ & \citet{Durech:2011ds} \\
 10 & 197.14 & 1.83  & STM   & \citet{Usui:2011kh} \\
 11 & 185.00 & 10.68 & NEATM & \citet{Masiero:2011jc} \\
 12 & 165.00 & 21.00 & LC+AO & \citet{Hanus:2013jh} \\
 13 & 195.64 & 5.44  & NEATM & \citet{Masiero:2014gu} \\
\hline \noalign {\smallskip}
   & 191.5 & 8.3  &       & Mean value$^*$ \\ % quadratic sum of errors or stddev of mean values??
   & 193 & 6     & ADAM (SPHERE only) & This paper \\
   & 196 & 6     & ADAM (all AO) & This paper \\
   & 198 & 4/2     & TPM & This paper \\
\hline \noalign {\smallskip}
\end{tabular}
\end{center} 
{\bf Note:} {}$^*$Using only values falling within 1-$\sigma$ of the average value of the full dataset.
\end{table*}
%-----------------------------Table End ------------------------------

%-----------------------------Figure Start-----------------------------
%-----------------------------Figure End ------------------------------

%-----------------------------Table Start-----------------------------
\begin{table*}[tbh!]
\footnotesize
\begin{center}
  \caption[]{\em{\small \rm Mass estimates of (6)~Hebe gathered from
      the literature. OD: orbital deflection, PE: planetary
      ephemeris.\\}} 
  \label{table:massestimates}
\begin{tabular}{rr@{\,$\pm$\,}r@{\,$\times$\,}lcl}
  \hline \noalign {\smallskip}
  & \multicolumn{3}{c}{Mass ($M$, kg)} & Method & Reference  \\
  \hline \noalign {\smallskip}
  1 & 1.37 & 0.44 & $10^{19}$ & OD & \citet{Michalak:2001ho} \\
  2 & 1.37 & 0.18 & $10^{19}$ & OD & \citet{Kochetova:2004ii} \\
  3 & 1.28 & 0.06 & $10^{19}$ & OD & \citet{Baer:2008jl} \\
  4 & 3.18 & 2.19 & $10^{17}$ & PE & \citet{Fienga:2009cm} \\
  5 & 9.07 & 0.91 & $10^{18}$ & PE & \citet{Folkner:2009wm} \\
  6 & 1.27 & 0.13 & $10^{19}$ & OD & \citet{Baer:2011er} \\
  7 & 1.41 & 0.24 & $10^{19}$ & PE & \citet{Fienga:2011dg} \\
  8 & 1.34 & 0.33 & $10^{19}$ & PE & \citet{Konopliv:2011fa} \\
  9 & 1.36 & 0.29 & $10^{19}$ & OD & \citet{Zielenbach:2011em} \\
 10 & 1.55 & 0.18 & $10^{19}$ & OD & \citet{Zielenbach:2011em} \\
 11 & 1.54 & 0.24 & $10^{19}$ & OD & \citet{Zielenbach:2011em} \\
 12 & 1.53 & 0.34 & $10^{19}$ & OD & \citet{Zielenbach:2011em} \\
 13 & 1.41 & 0.17 & $10^{19}$ & PE & \citet{Fienga:2013wd} \\
 14 & 8.39 & 1.95 & $10^{18}$ & PE & \citet{Kuchynka:2013eu} \\
 15 & 8.06 & 0.91 & $10^{18}$ & PE & \citet{Pitjeva:2013fb} \\
 16 & 8.95 & 0.60 & $10^{18}$ & OD & \citet{Goffin:2014in} \\
 17 & 1.21 & 0.08 & $10^{19}$ & OD & \citet{Kochetova:2014ia} \\
 18 & 1.86 & 0.13 & $10^{19}$ & PE & \citet{Fienga:2014ux} \\
 19 & 1.58 & 0.19 & $10^{19}$ & PE & Fienga (private comm.) \\
 \hline \noalign {\smallskip}
    & 1.31 & 0.24 & $10^{19}$ &  & Mean value$^*$ \\
 \hline \noalign {\smallskip}
\end{tabular}
\end{center} 
{\bf Note:} {}$^*$Using only values falling within 1-$\sigma$ of the average value of the full dataset.
\end{table*}
%-----------------------------Table End ------------------------------

%-----------------------------Figure Start-----------------------------
%-----------------------------Figure End ------------------------------

\section{Thermophysical model \label{sec:tpm}}

\begin{figure}[h!]
\centering
\includegraphics[angle=90, width=\linewidth, trim=2cm 3cm 2cm 3cm, clip]{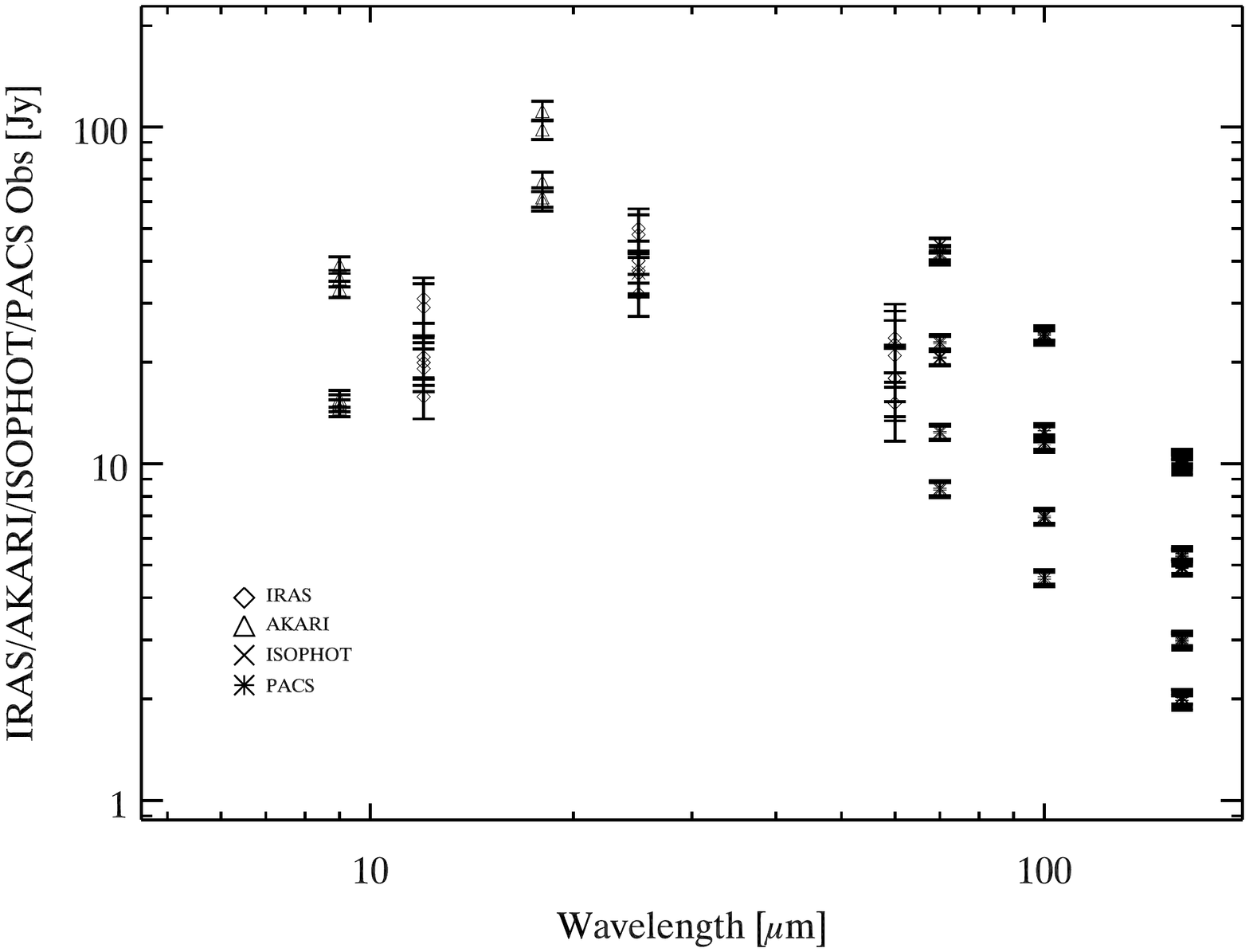}

 \caption[ ]{Thermal flux measurements of (6)-Hebe used 
 for the thermophysical modeling.  
 From IRAS (12, 25, 60, 100\,$\mu$m \citealt{Tedesco:2002fs}),
  AKARI-IRC (9, 18\,$\mu$m \citealt{Usui:2011kh}), 
  ISO-ISOPHOT (25\,$\mu$m, \citealt{Lagerros:1999ku}), and
  Herschel-PACS (70, 100, 160\,$\mu$m, M\"uller et al., in prep).}
\label{fig:hebe-thermal}
\end{figure}

  The thermophysical model (TPM) used in this work predicts for a given 
  set of parameters, including the volume-equivalent
  diameter $D$, albedo p$_{\rm v}$, surface roughness $\bar \theta$,
  and thermal inertia $\Gamma$, a flux that can be compared
  to the observed flux. The input parameters can then be optimized by minimizing
  the reduced $\chi^2$ between the model and observations.
  Thermal measurements of Hebe used in the modeling procedure are plotted in
  Figure\,\ref{fig:hebe-thermal}.

\begin{figure}[h!]
\centering
\includegraphics[angle=90, width=\linewidth, trim=0cm 0cm 0cm 0cm, clip]{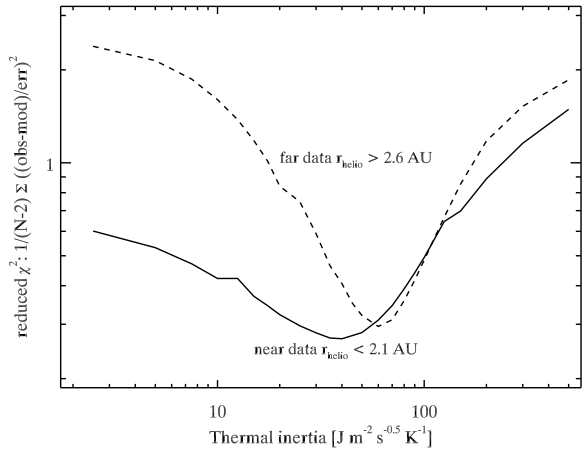}

 \caption[ ]{Thermal inertia of (6)~Hebe derived from the thermophysical modeling. 
 The overall preferred solution (lower reduced $\chi^2$) is $\sim$60~J\,m$^{-2}$\,s$^{-0.5}$K\,$^{-1}$ 
  for data acquired at heliocentric distance r$<$2.1~AU and $\sim$40~J\,m$^{-2}$\,s$^{-0.5}$K\,$^{-1}$ for data 
  taken at r$>$2.6 AU. While this might be indicative of changing thermal inertia with temperature, one should
  be extremely cautious when interpreting this result, as a range of solutions cannot be ruled out based on the 
  $\chi^2$ values presented here.}
\label{fig:hebe-thermalinertia}
\end{figure}

\begin{figure}[h!]
\centering
\includegraphics[angle=90, width=\linewidth, trim=2cm 3cm 2cm 3cm, clip]{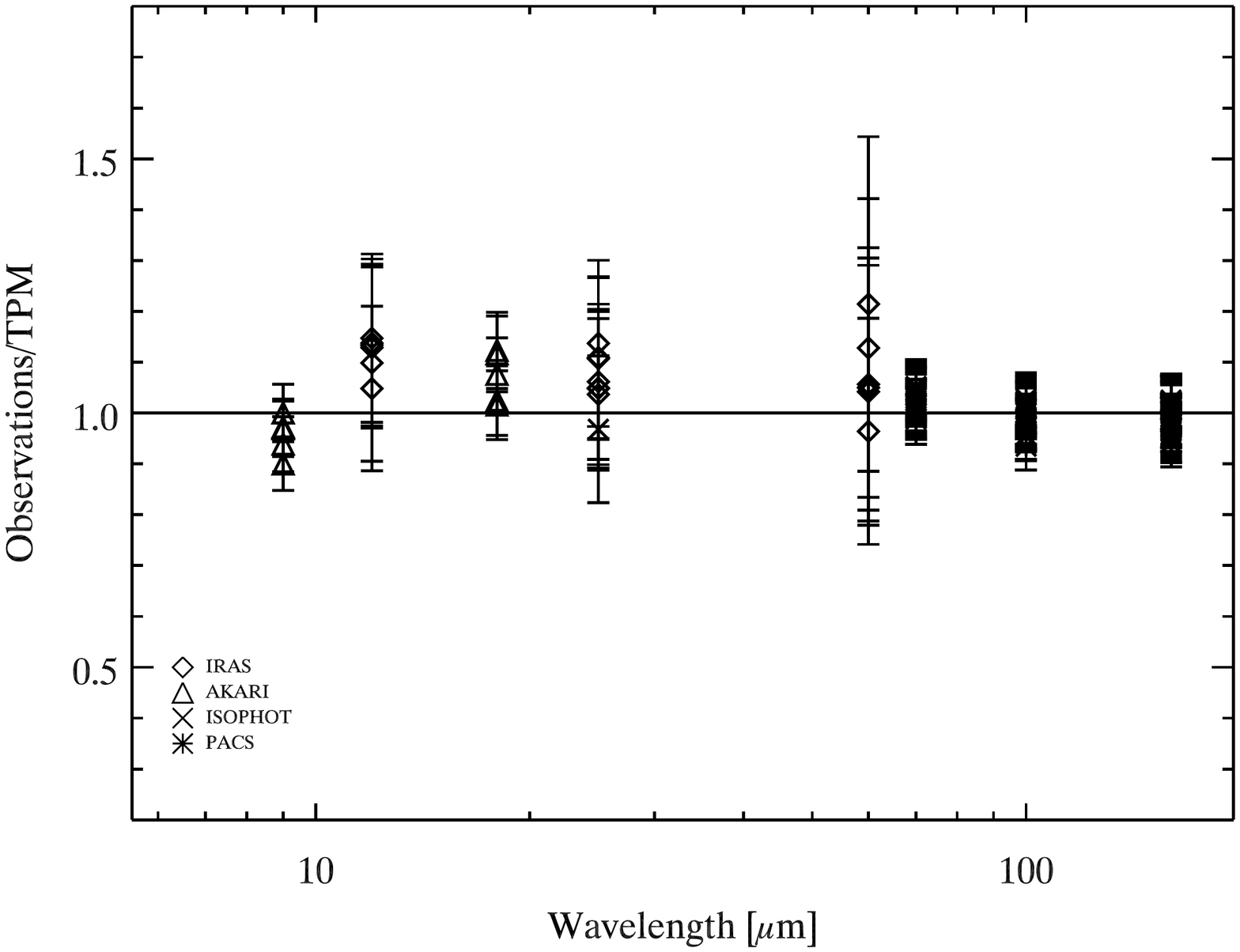}

 \caption[ ]{Observation-to-model flux ratios as a function of wavelengths, based on color-corrected
     mono-chromatic flux densities and the corresponding TPM flux predictions.}
\label{fig:hebe-thermal2}
\end{figure}

  Here, a solution was derived simultaneously for $\Gamma$, $D$ and p$_{\rm v}$
  for a range of different $\bar \theta$ 
  knowing Hebe's absolute magnitude H and magnitude slope G. 
  Different emissivity models, including constant e=0.9 and
  wavelength-dependent emissivities, were tested. We adopted  
  the emissivity model for large main-belt asteroids of \citet{Mueller:1998to} 
  which was found to provide the most satisfactory results (lower $\chi^2$).
  Finally, best-fit solutions were found for significant surface
  roughness and $\Gamma$ values ranging from 20 to
  100~J\,m$^{-2}$\,s$^{-0.5}$K\,$^{-1}$ (Figure\,\ref{fig:hebe-thermalinertia}).
  The resulting observation-to-model flux ratios are shown at Figure\,\ref{fig:hebe-thermal2}.

  %Remarkably, the best-fitting solution for $\Gamma$
  %drops from $\sim$60~J\,m$^{-2}$\,s$^{-0.5}$K\,$^{-1}$ 
  %when only considering data taken at r$<$2.1~AU to
  %$\sim$40~J\,m$^{-2}$\,s$^{-0.5}$K\,$^{-1}$ for data 
  %taken at r$>$2.6 AU, in perfect agreement with the temperature-dependancy 
  %of thermal conductivity and specific heat (Figure\,\ref{fig:hebe-thermalinertia}).

%-----------------------------Figure Start-----------------------------
%-----------------------------Figure End ------------------------------

%-----------------------------Figure Start-----------------------------
%-----------------------------Figure End ------------------------------

%-----------------------------Figure Start-----------------------------
%-----------------------------Figure End ------------------------------

We further used a well established method \citep{Gundlach:2013kn} to determine 
  the grain size of the surface regolith of  Hebe. 
  The method consists in estimating the heat conductivity of the surface material derived 
  from the thermal inertia measurements and then to compare the values with calculations 
  of the heat conductivity of a model regolith for distinct volume-filling factors of the regolith grains. 
  The thermal inertia value and the surface temperature of these bodies are two input parameters 
  for the method. 
  First of all the thermal inertia $\Gamma$ is used to calculate the conductivity $\kappa$ using:
\begin{equation}
\kappa = \frac{\Gamma^2}{\phi \rho c},
\label{E:kappaAst}
\end{equation}
where $c$ is the specific heat capacity, $\rho$ the material density, and $\phi$ the regolith volume-filling factor, which is typically unknown. So, this last parameter is varied between 0.6 (close to the densest packing of equal-sized particles) and 0.1 (extremely fluffy packing, plausible only for small regolith particles) with $\Delta \phi$=0.1, while here we take values for $\rho$ and $c$ typical of H5 ordinary chondrites from \cite{Opeil:2010hq}. We estimate Hebe's temperature to be 230 K and 180 K for the thermal inertia determination at 1.94 and 2.87 AU, respectively. 

\begin{figure}[h!!!!!]
\includegraphics[width=\linewidth]{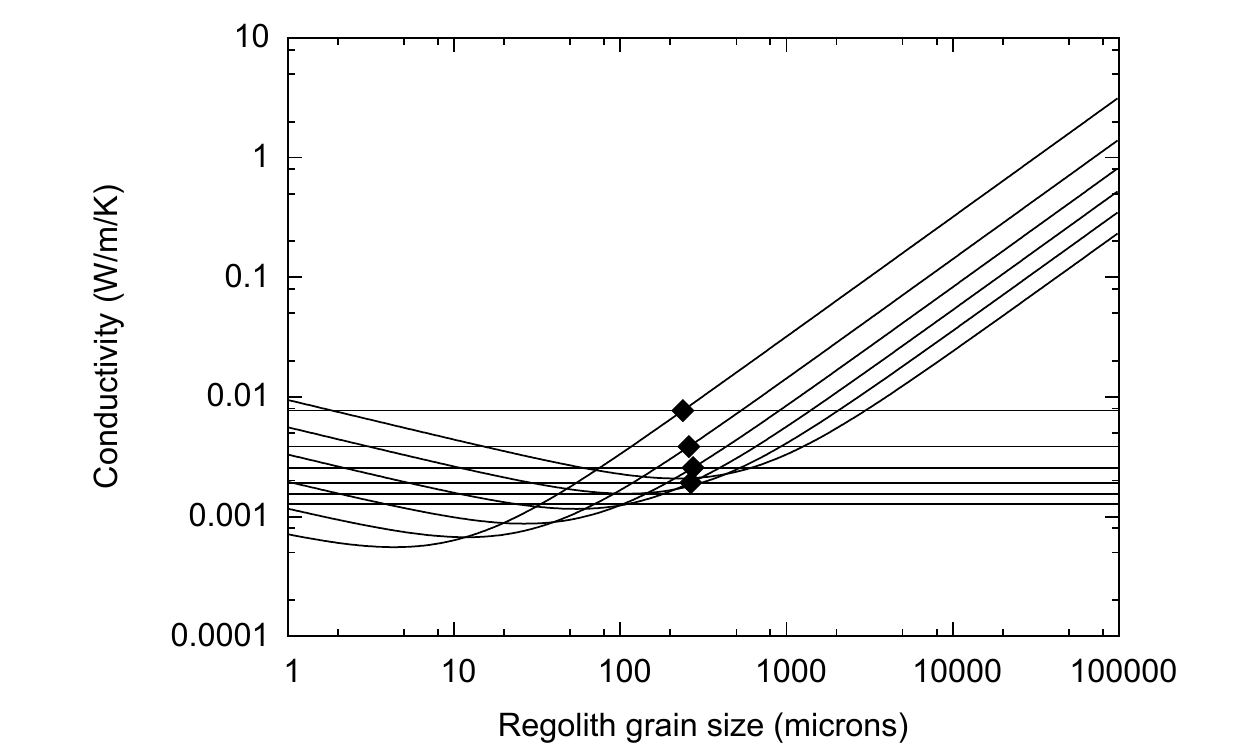}
\caption{Hebe's regolith grain size. Horizontal lines indicate the derived values of the heat conductivity, following Eq.~\ref{E:kappaAst}, for the different volume-filling factors of the material and for a thermal-inertia value of 40~J\,m$^{-2}$\,s$^{-0.5}$K\,$^{-1}$ and a surface temperature of 180 K. The curves represent the thermal conductivity of a regolith with thermophysical properties of a H5 meteorite as from \cite{Opeil:2010hq} as a function of the regolith grain size. The intersection of the curves with the horizontal lines gives the grain size of the regolith.}
\label{fig:granSizeLowGamma}
\end{figure}

\begin{figure}[h!!]
\includegraphics[width=\linewidth]{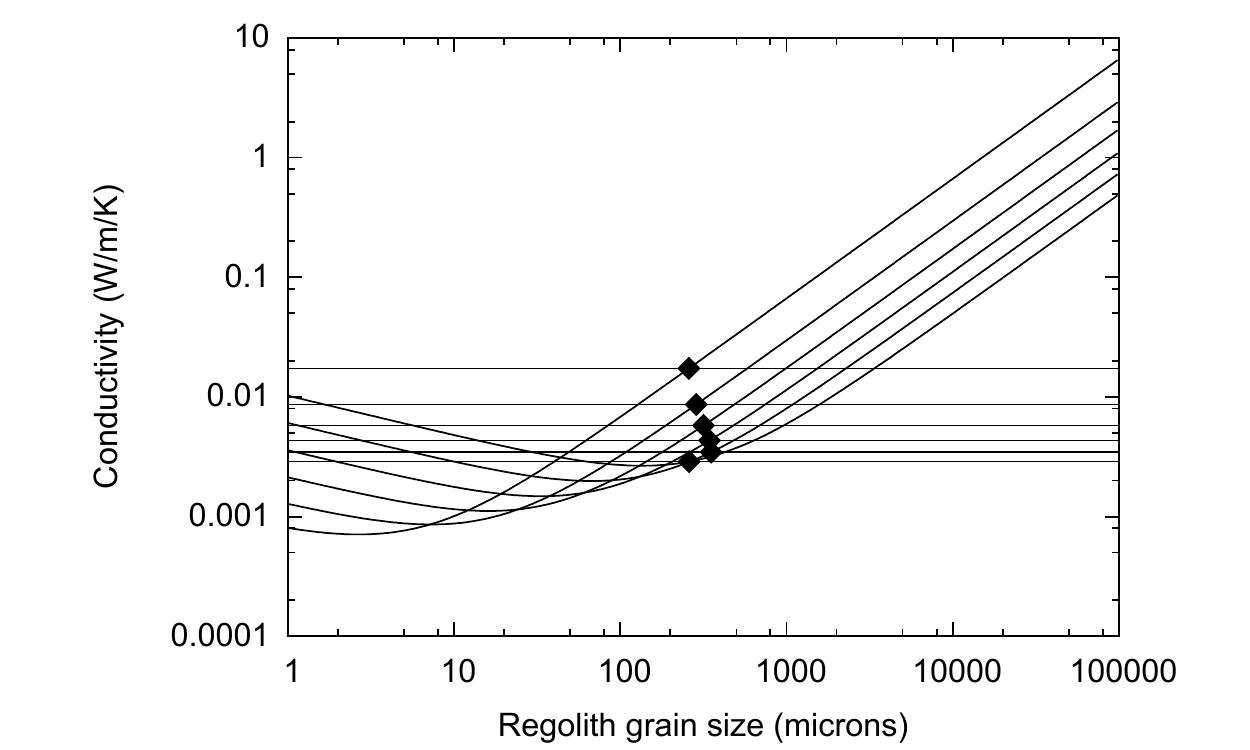}
\caption{Same as Fig.~\ref{fig:granSizeLowGamma} but showing the regolith grain size for the a thermal inertia of 60~J\,m$^{-2}$\,s$^{-0.5}$K\,$^{-1}$ and a surface temperature of 230 K.}
\label{fig:granSizeHighGamma}
\end{figure}

By doing so, we find a typical grain size of 0.2--0.3\,mm (Figures~\ref{fig:granSizeLowGamma} and \ref{fig:granSizeHighGamma}).

\end{document}